\def \tr {{\rm tr}\,}
\documentclass[twocolumn,showpacs,preprintnumbers,amsmath,amssymb]{revtex4}
\usepackage{epsf}


\begin{document}
\title{Constraint-preserving boundary conditions in the Z4 Numerical
Relativity formalism}
\author{C~Bona$^1$, T~Ledvinka$^2$, C~Palenzuela-Luque$^3$ and
M.~\v Z\' a\v cek$^2$}

\affiliation{ $^1$ Departament de Fisica, Universitat de les Illes
Balears, Palma de Mallorca, Spain \\
$^2$ Institute of Theoretical Physics, Charles University, Prague,
Czech Republic \\
$^3$ Department of Physics and Astronomy, Louisiana State
University, Louisiana, USA }

\begin{abstract}
The constraint-preserving approach is discussed in parallel with
other recent developments with the goal of providing consistent
boundary conditions for Numerical Relativity simulations. The case
of the first order version of the Z4 system is considered, and
constraint-preserving boundary conditions of the Sommerfeld type
are provided. The stability of the proposed boundary conditions is
related with the choices of the ordering parameter. This
relationship is explored numerically and some values of the
ordering parameter are shown to provide stable boundary conditions
in the absence of corners and edges. Maximally dissipative
boundary conditions are also implemented. In this case, a wider
range of values of the ordering parameter is allowed, which is
shown numerically to provide stable boundary conditions even in
the presence of corners and edges.
\end{abstract}

\maketitle

\section{Introduction}

The relevance of the initial boundary value problem (IBVP) for
Numerical Relativity has been pointed out many times since the
ground-breaking work of Stewart~\cite{Stewart}.

The origin of the problem is the well-known fact that Einstein's
equations
\begin{equation}\label{Einsteins}
    G^{\mu\nu} = 8\pi~T^{\mu\nu},
\end{equation}
when interpreted as second-order field equations for the metric
components $g_{\mu\nu}$, provide only six evolution equations for
the space components $g_{ij}$ whereas the remaining four
Einstein's equations
\begin{equation}\label{Constraints}
    G^{\mu 0} = 8\pi~T^{\mu 0}
\end{equation}
are just second-order constraints on $g_{ij}$~\cite{ADM}. The
evolution equations are then a reduction of the full Einstein's
system. Notice that this reduction is not uniquely defined, as far
as one can add in multiples of the constraints to the original
evolution system: this freedom is at the root of the diversity of
the proposed evolution formalisms (for a review, see
Ref.~\cite{Reula}).

The current approach in Numerical Relativity is to use one of
these reductions of Einstein's field equations, plus four
coordinate conditions, as the main evolution system in order to
compute the full set of metric components. This is the
unconstrained, or free evolution~\cite{Cent80}, approach in which
the constraint equations (\ref{Constraints}) are mainly used for
monitoring the accuracy of the simulations. As a consequence, the
evolution system has an extended space of solutions which
contains, in addition to the true ones, constraint-violating
solutions that do not verify the full set of Einstein's equations.

This poses the question of which are the requirements in order to
get, in the free evolution approach, true Einstein's solutions in
a consistent and stable way, avoiding any drifting towards
extended, constraint-violating solutions~\cite{Stewart}. The key
point is to analyze the subsidiary system
\begin{equation}\label{subsidiary}
    \nabla_\nu ( G^{\mu\nu} - 8\pi~T^{\mu\nu} ) = 0,
\end{equation}
which follows from the contracted Bianchi identities and the
conservation of the stress-energy tensor. As it is well known, the
subsidiary system ensures that the constraints (\ref{Constraints})
are first integrals of the main evolution system. But it can also
be interpreted as providing evolution equations for the
constraints deviations. Of course, the same is true for any
variant obtained by combining (\ref{subsidiary}) with space
derivatives of either evolution or constraint equations: this
freedom can be easily used in order to get a variant of the
subsidiary system (\ref{subsidiary}) with a strongly
hyperbolic~\cite{KL89}, even symmetric hyperbolic, principal part:
\begin{eqnarray}
    \partial_0 ( G^{00} - 8\pi~T^{00} ) +
    \partial_k ( G^{0k} - 8\pi~T^{0k} )&=& \cdots \\
    \partial_0 ( G_{k}^{~0} - 8\pi~T_{k}^{~0} ) +
    \partial_k ( G^{00} - 8\pi~T^{00} )&=& \cdots
\end{eqnarray}
(normal coordinates).

In the case of the pure initial value problem (IVP), the question
about the consistency and stability for free-evolution
true-Einstein's solutions has been given a precise answer in
Ref.~\cite{Stewart}:
\begin{itemize}
    \item Consistency: The initial data must verify the constraint
equations.
    \item Stability: The principal part of both the main evolution
system and the subsidiary system must be strongly hyperbolic.
\end{itemize}
Our former considerations suggest that the strong hyperbolicity
requirement on the subsidiary system is always fulfilled in the
cases in which the only constraints are the energy-momentum ones
(\ref{Constraints}).

In the case of the IBVP, which is the main subject of this paper,
a number of developments are currently under way. Let us briefly
summarize the main ones:

\subsubsection{Constraint-preserving boundary conditions.}

The original work in Ref.~\cite{Stewart} was focused on the
Frittelli-Reula evolution system~\cite{FR94,FR96}. More
recently~\cite{CLT02, CPRST03, CS03,Calabrese04}, the
constraint-preserving boundaries approach has been extended to
other symmetric-hyperbolic systems of the Kidder-Scheel-Teukolsky
(KST) type~\cite{KST01,ST02}. The full programme consists in the
following steps:
\begin{itemize}
    \item Writing down the subsidiary system (it can be
    (\ref{subsidiary}) or any variant of it) as a first order
    evolution system for constraint deviations, with symmetric
    hyperbolic principal part.
    \item Providing algebraic (Dirichlet) boundary conditions for the
    incoming modes of the subsidiary system in such a way that the
    total amount of constraint deviations (as measured with a suitable
    energy estimate) keeps bounded.
    \item Interpreting these algebraic boundary conditions of the
    subsidiary system as differential (Neumann) boundary
    conditions for the constraint-related incoming modes of the main
    evolution system.
    \item Completing the resulting subset of boundary conditions by
    adding suitable conditions for the remaining incoming modes.
    \item Checking the stability of the final set of the main system
    boundary conditions. In Ref.~\cite{Stewart}, the
    Majda-Osher theory~\cite{MO75} is applied and the uniform Kreiss
    condition~\cite{KL89} is obtained as a result.
\end{itemize}

\subsubsection{Einstein's boundary conditions}

An alternative approach has been proposed by Frittelli and
G\'{o}mez~\cite{FG03a, FG03b,FG04a, FG04b}. For a better
understanding, let us start by writing down the constraint
equations (\ref{Constraints}) in a covariant form
\begin{equation}\label{covconstraints}
        n_\nu (G^{\mu\nu} - 8\pi~T^{\mu\nu}) = 0\,,
\end{equation}
where $n_\nu$ is the normal to the constant-time hypersurfaces. In
local adapted coordinates we have
\begin{equation}\label{timenormal}
    n_\nu = \alpha~\delta_\nu^{~0}\,,
\end{equation}
so that the original form (\ref{Constraints}) is recovered. The
absence of second time derivatives in (\ref{Constraints}) can now
be rephrased as the absence of second derivatives normal to the
spacelike constant-time hypersurfaces.

Now let us invoke the general covariance of Einstein's theory. We
will realize that the same kind of result must hold for other kind
of hypersurfaces, not just the spacelike ones. We can then start
again from the covariant form (\ref{covconstraints}) but with
$n_\nu$ being now the normal to any timelike hypersurface, like
the ones corresponding to the boundaries of our computational
domain. In local adapted coordinates, we could take for instance
\begin{equation}\label{spacenormal}
    n_\nu \sim \delta_\nu^{~z}\,,
\end{equation}
so that no second derivatives normal to the constant-z
hypersurfaces would appear in (\ref{covconstraints}).

If the main evolution system is written in first order form, the
absence of second order normal derivatives ensures that four
combinations of the first-order dynamical fields can be
consistently computed at the boundaries without any recourse to
outside information. The Fritelli-G\'{o}mez idea is to use these
combinations in order to get consistent boundary conditions for a
subset of four incoming modes (Einstein's boundaries). Of course,
suitable conditions for the remaining incoming modes must also be
provided and the stability of the full set must be checked.

\subsubsection{Harmonic coordinates}

Still another approach is due to Szil\'{a}gyi, Winicour and
coworkers~\cite{SGBW00, SSW02, SW03, BSW04}. Their formulation
looks quite different from the preceding ones, so that we will
need to rephrase some statements in order to point out the
underlying similarities.

For instance, instead of the reduction of Einstein's equations, we
will consider the equivalent extension of the solution space. In
the harmonic coordinates approach, this extension is achieved by
writing down the principal part of Einstein's equations as a set
of generalized wave equations with some extra terms (de
Donder-Fock decomposition~\cite{DeDo21,Fock59}) and then getting
rid of these extra terms by requiring the four spacetime
coordinates to be harmonic functions, that is
\begin{equation}\label{4harmonic}
    \Box~x^\mu = 0
\end{equation}
($x^\mu$ is considered as a set of four scalar functions here).

Let us compare now the Harmonic Coordinates approach with the
preceding ones:
\begin{itemize}
    \item The resulting (relaxed) system is used as the main
evolution system for the full set of metric components. By
construction, its principal part amounts to a set of wave
equations, so that symmetric hyperbolicity is ensured.
    \item True Einstein's solutions are recovered only when
imposing the coordinate conditions (\ref{4harmonic}), which play
here the role of the constraints. The extra principal terms that
were suppressed from the original Einstein's system contained
first derivatives of these coordinate constraints (second
derivatives of the metric components)~\cite{SW03}.
    \item The subsidiary system can be again obtained from
(\ref{subsidiary}). Terms containing the main evolution system
equations, or their derivatives, vanish separately so that only
the contribution of the extra principal terms remains. Notice that
the resulting subsidiary system is of second order in the
coordinate constraints (\ref{4harmonic}), in contrast with the
former approaches.
    \item The principal part of the (second order) subsidiary
    system is symmetric hyperbolic: it amounts again to a set of
    wave equations on the coordinate constraints (\ref{4harmonic}).
\end{itemize}

In Ref.~\cite{SW03}, a boundary condition derived from the local
reflection symmetry requirement is analyzed. Constraint
preservation is explicitly shown and a theorem of
Secchi~\cite{Secchi96} is used in order to show that the resulting
IBVP is well posed. This theoretical result is checked by means of
the numerical robust-stability test~\cite{Mexico1}, which is
adapted so that reflection boundary conditions are applied along
just one space axis, while keeping periodic boundary conditions
along the other two.

Due to the limited use of reflection symmetry boundary conditions
in practical applications, a proposal is made~\cite{BSW04} for
extending these results to boundary conditions of the Sommerfeld
type, as it was done previously in a different
framework~\cite{FN99}.

\subsubsection{The Z4 case}\label{Z4case}

The Z4 approach~\cite{Z4,Z48} uses an extra dynamical four-vector,
along the track of previous formulations which contained extra
dynamical quantities~\cite{BM92,BM95,SN95,BS99}. It has strong
similarities with the harmonic coordinates approach, while
providing much more flexibility regarding the gauge choices.
\begin{itemize}
    \item The main evolution system is obtained by modifying
Einstein's equations with the help of the extra four-vector
$Z^\mu$, namely
\begin{equation}\label{EinsteinZ4}
    G_{\mu\nu} + \nabla_{\mu} Z_{\nu} + \nabla_{\nu} Z_{\mu} -
    ( \nabla_{\rho} Z^{\rho} )~g_{\mu\nu} = 8~\pi~T_{\mu\nu}\,.
\end{equation}
This system  provides ten evolution equations for the set formed
by the six space components of the metric plus the four $Z^\mu$
components. A first order version can be easily obtained which,
when supplemented with suitable gauge conditions, has been shown
to have a strongly hyperbolic principal part~\cite{Z48}.
    \item True Einstein's solutions can be recovered by requiring
the vanishing of the extra four-vector
\begin{equation}\label{Zis0}
    Z^{\mu} = 0\,,
\end{equation}
so that this condition can be considered as a set of four
algebraic constraints. Notice that the main evolution system
(\ref{EinsteinZ4}) is of a mixed type: it contains second order
derivatives of the metric, but only first order derivatives of the
extra four-vector.
    \item The subsidiary system can be obtained from the covariant
divergence of the main system (\ref{EinsteinZ4}): it will be of
third order in the metric and of second order on the constraint
variables $Z_\mu$. Allowing for (\ref{subsidiary}), the Einstein's
tensor contribution vanishes separately, so that only the
contribution of the extra terms remains, namely
\begin{equation}\label{divZ4}
    \nabla_{\nu}~[~\nabla^{\mu} Z^{\nu} + \nabla^{\nu} Z^{\mu} -
    ( \nabla_{\rho} Z^{\rho} )~g^{\mu\nu}~]~ = 0~,
\end{equation}
which can be also expressed in the equivalent form~\cite{Z48}
\begin{equation}\label{WaveZis0}
  \Box~ Z_{\mu} + R_{\mu\nu} Z^\nu = 0~.
\end{equation}
    \item It follows from (\ref{WaveZis0}) that the subsidiary
system, when considered as a second order system for the algebraic
constraints deviations $Z_\mu$, has a symmetric hyperbolic
principal part: it amounts here again to an uncoupled set of wave
equations.
\end{itemize}

The fact that the subsidiary system is of second order means that
the vanishing of both $Z_\mu$ and its first time derivative must
be imposed on the initial data if one wants to ensure \textit{a
priori} that the resulting solution will be a true Einstein's one.
This amounts to impose the usual energy and momentum constraints
(\ref{Constraints}) on the initial data hypersurface, so that it
can seem that the Z4 formalism is not being of much help. But on
the other side, if one is checking a given solution \textit{a
posteriori}, the vanishing of $Z\mu$ in a given spacetime domain
ensures that the same is true for its derivatives, so that this
solution is necessarily a true Einstein's one. This is why one can
monitor constraint violations by looking just at the values of
$Z_\mu$ and, more important, this is why one can devise
constraint-preserving strategies by aiming at the vanishing of
$Z_\mu$. Here is where the Z4 formalism shows its main advantages.

In what follows, we will consider the fully first-order version of
the Z4 system~\cite{Z48}, as summarized in Section 2. In Section
3, we will apply the constraint-preserving boundary conditions
programme, obtaining as a result conditions of the Sommerfeld type
for the main evolution system. The stability of these conditions
is studied in Section 4, including the use of the robust-stability
numerical test. As a result, our Sommerfeld-like conditions will
be shown to behave in the same way as the reflection symmetry ones
proposed in Refs.~\cite{SW03,BSW04}.

\section{First order Z4 system}

The general-covariant equations (\ref{EinsteinZ4}) can be written
in the equivalent 3+1 form \cite{Z4}
\begin{eqnarray}
\label{dtgamma}
  (\partial_t -{\cal L}_{\beta})~ \gamma_{ij}
  &=& - {2~\alpha}~K_{ij}
\\
\label{dtK}
   (\partial_t - {\cal L}_{\beta})~K_{ij} &=& -\nabla_i\alpha_j
    + \alpha~   [{}^{(3)}\!R_{ij}
    + \nabla_i Z_j+\nabla_j Z_i
\nonumber \\
    &~&-~ 2\,K^2_{ij}+(\tr K-2\Theta)~K_{ij}
\nonumber \\
    &~&- ~S_{ij}+\frac{1}{2}\,(\tr S - \tau)~\gamma_{ij}~]
\\
\label{dtTheta} (\partial_t -{\cal L}_{\beta})~\Theta &=&
\frac{\alpha}{2}~
 [{}^{(3)}\!R + 2~ \nabla_k Z^k + (\tr K - 2~ \Theta)~\tr K
\nonumber \\
&~&- ~\tr(K^2) - 2\tau~] - Z^k {\alpha}_k
\\
\label{dtZ}
 (\partial_t -{\cal L}_{\beta})~Z_i &=& \alpha~ [~\nabla_j\,({K_i}^j
  -{\delta_i}^j \tr K) + \partial_i \Theta
\nonumber \\
  &~&- ~2\,{K_i}^j~ Z_j - S_i~] - \Theta\,{\alpha}_i
\end{eqnarray}
where we have noted
\begin{eqnarray}\label{tauSdef}\nonumber
  \tau \equiv  8 \pi  \alpha^2~ T^{00}\,,~
  S_i &\equiv&  8 \pi \alpha ~ T^0_{~i}\,,~
  S_{ij} \equiv 8 \pi ~T_{ij}\,,~  \\
  \Theta \equiv  \alpha ~ Z^0\,,&~& \alpha_i \equiv
  \partial_i~\alpha\,.
\end{eqnarray}

In the form (\ref{dtgamma}-\ref{dtZ}), it is evident that the Z4
evolution system is fully relaxed: it consists only of evolution
equations. The original constraints (\ref{Zis0}), which can be
translated into
\begin{equation}\label{ZThis0}
  \Theta~=~0,\qquad Z_i~=~0,
\end{equation}
are algebraic so that the full set of field equations
(\ref{EinsteinZ4}) is actually used during evolution, like in the
harmonic coordinates case.

But now we have not to impose the harmonic coordinate conditions
(\ref{4harmonic}). We will consider instead a wider class of gauge
conditions, in which the time slicing will be of the
form~\cite{Z48}
\begin{equation}\label{dtAlpha}
 (\partial_t -{\cal L}_{\beta})~\ln \alpha = -~f\alpha~({\tr} K
 - m \Theta)
\end{equation}
(generalized harmonic slicing). Although more general cases can be
considered~\cite{BP04}, we will use here normal coordinates (zero
shift) for simplicity.

A first order version of the Z4 evolution system
(\ref{dtgamma}-\ref{dtZ}) can be obtained by introducing the first
space derivatives
\begin{equation}\label{AkDkij}
 A_k~\equiv~\alpha_k/\alpha,~~D_{kij}~\equiv~\frac{1}{2}~\partial_k \gamma_{ij}
\end{equation}
as independent dynamical quantities, so that the full set of
dynamical fields can be given by
\begin{equation}\label{uvector}
 \mathbf{u} ~ = ~ \{\alpha,~\gamma_{ij},~ K_{ij},~ A_k,~D_{kij},~\Theta,~Z_k \}
\end{equation}
(38 independent fields).

Of course, one must provide evolution equations for the new
quantities (\ref{AkDkij}): the simplest way is to take
\begin{eqnarray}\label{dtA}
 \partial_t A_k~&+&~\partial_k [~ f \alpha~( {\tr}K - m \Theta) ~]~=~0
\\\label{dtD}
 \partial_t D_{kij}~&+&~\partial_k [~\alpha~K_{ij}~]~=~0~.
\end{eqnarray}
Notice that one could add to (\ref{dtA}, \ref{dtD}) a number of
terms involving first derivatives of either $\Theta$ or $Z_k$.
This would amount to introduce coupling terms with either the
Energy or the Momentum constraints, as in the KST
system~\cite{KST01, ST02}, each one with its own free parameter.

We have chosen instead to keep the simplest form (\ref{dtA},
\ref{dtD}) because the first order constraints (\ref{AkDkij})
evolve in a trivial way, that is
\begin{eqnarray}\label{dtAbis}
 \partial_t [~ A_k~-~\partial_k ~ln\, \alpha ~]~&=&~0
\\\label{dtDbis}
 \partial_t [~ D_{kij}~-~\partial_k~ \gamma_{ij}~]~&=&~0~,
\end{eqnarray}
so that the relationship between the first and the second order
versions of the evolution system is more transparent. We are
losing in this way the possibility of playing with a number of
extra free parameters.

Care must be taken, however, when expressing the Ricci tensor
${}^{(3)}\!R_{ij}$ in (\ref{dtK}) in terms of the derivatives of
$D_{kij}$, because as far as the definitions (\ref{AkDkij}) are no
longer enforced, the identities
\begin{equation}\label{dDdD}
 C_{kl} \equiv \partial_{[k}~A_{l]} = 0\,\qquad
 C_{klij} \equiv \partial_{[k}~D_{l]ij} = 0
\end{equation}
can not be taken for granted in first order systems. As a
consequence of these ordering ambiguities, the principal part of
the evolution equation (\ref{dtK}) leads to a one-parameter family
of non-equivalent first-order versions, namely
\begin{equation}\label{PrincipalK}
 \partial_t K_{ij} ~+~\partial_k~[~\alpha~\lambda^k_{ij}~]~=~...~
\end{equation}
where
\begin{eqnarray}\label{deflambda}
 \lambda^k_{ij} &=& {D^k}_{ij}
   -{\frac{1+\zeta}{2}}~ (D_{ij}^{~~k}+D_{ji}^{~~k}
   -\delta^k_i E_j-\delta^k_j E_i)
\nonumber \\
 &+& \frac{1}{2}\, \delta^k_i(A_j-D_j+2V_j)
 + \frac{1}{2}\, \delta^k_j(A_i-D_i+2V_i)\qquad
\end{eqnarray}
and we have noted
\begin{equation}\label{defVk}
    D_i \equiv \gamma^{rs}D_{irs}\,,~ E_i \equiv
    \gamma^{rs}D_{rsi}\,,~
  V_k \equiv D_k-E_k-Z_k\,.
\end{equation}
Notice that the parameter choice $\zeta = +1$ corresponds to the
standard Ricci decomposition
\begin{equation}\label{Def3R}
{}^{(3)}\!R_{ij}~=~\partial_k~{\Gamma^k}_{ij}-\partial_i~{\Gamma^k}_{kj}
+{\Gamma^r}_{rk}{\Gamma^k}_{ij}-{\Gamma^k}_{ri}{\Gamma^r}_{kj}
\end{equation}
whereas the opposite choice $\zeta = -1$ corresponds to the
de~Donder-Fock \cite{DeDo21,Fock59} decomposition
\begin{eqnarray}\label{Def3dDF}
{}^{(3)}\!R_{ij}&=&-\partial_k~{D^k}_{ij}+\partial_{(i}~{\Gamma_{j)k}}^{k}
- 2 {D_r}^{rk} D_{kij} \nonumber \\
&+& 4 {D^{rs}}_i D_{rsj} - {\Gamma_{irs}}
{\Gamma_j}^{rs}-{\Gamma_{rij}} {\Gamma^{rk}}_k
\end{eqnarray}
which is most commonly used in Numerical Relativity formalisms.
The ordering ambiguities do not affect to the principal part of
eq. (\ref{dtTheta}), namely
\begin{equation}
 \partial_t~\Theta + \partial_k ~[~\alpha~ V^k~] = ...
\end{equation}

The resulting first order system has been shown to be strongly
hyperbolic~\cite{Z48} provided that the first gauge parameter $f$
is greater than zero. In the harmonic slicing case ($f=1$), the
second gauge parameter is fixed ($m=2$). The full list of
eigenvectors is given in Appendix A.

\section{Constraint-preserving boundary conditions}

We have seen in the Introduction that the simple equation
(\ref{WaveZis0}) provides the subsidiary system for the deviations
of the algebraic constraints (\ref{Zis0}). This would be the whole
story if we were planning to use the second order version
(\ref{EinsteinZ4}) of the evolution system. But we prefer to focus
here in the first-order-in-space version, as described in the
previous section. The reason is that the mathematical theory of
first order systems seems to be more developed, both at the
continuum and at the discrete level, so more powerful tools are
available: Energy methods, Total-Variation-Diminishing algorithms
and so on (see for instance Refs.~\cite{KL89, GKO95}).

There is a price to pay for this. We have found in the previous
Section new constraints, like (\ref{dDdD}), arising from ordering
ambiguities in the space derivatives. The ordering parameter
$\zeta$ appeared precisely from the coupling of these ordering
constraints with the evolution system. The original subsidiary
system (\ref{WaveZis0}) must then be extended in order to include
both these coupling terms and the evolution of the ordering
constraints themselves.

\setcounter{subsubsection}{0}
\subsubsection{First-order subsidiary system.}

The easiest way of obtaining the full subsidiary system in the
first-order case is just by computing the time derivative of the
full Z4 first-order system. We give here (the principal part of)
the resulting subsidiary system
\begin{eqnarray}
  \partial_t~C_{kl} &=& 0
   \label{Asub} \\
  \partial_t~C_{klij} &=& 0
  \label{Dsub}\\
  1/\alpha^2~\partial^2_{tt}~\Theta - \triangle~\Theta&=& \cdots
  \label{Thetasub}\\ \nonumber
  1/\alpha^2~\partial^2_{tt}~Z_i - \triangle~Z_i &=&
  \gamma^{kl}~\partial_k~[~C_{il}
   +~\gamma^{rs}~(~C_{ilrs} \\
   + ~(\zeta-1)~C_{rlsi}
   &+& ~(\zeta+1)~C_{risl}~)~] + \cdots \qquad
  \label{Zsub}
\end{eqnarray}
(the dots stand for non-principal terms).

The subsidiary system (\ref{Asub} - \ref{Zsub}) can be put in
first order form in the usual way, by considering the first
derivatives of ($\Theta$, $Z_i$) as new independent variables. The
following evolution conditions
\begin{eqnarray}
  \partial_t~ (\partial_k\Theta)
  - \partial_k~ [~\partial_t\Theta~] &=& 0 \label{dtdtTheta}\\
  \partial_t~ (\partial_k Z_i)
  - \partial_k~ [~\partial_t Z_i~] &=& 0
\label{dtdtz}
\end{eqnarray}
could be added then to complete (the first order version of) the
subsidiary system.

Notice that the evolution equations (\ref{Asub}, \ref{Dsub}) for
the ordering constraints are trivial. This means that the ordering
constraints themselves are eigenfields of the full subsidiary
system (\ref{Asub} - \ref{dtdtz}) with zero characteristic speed.
Moreover, the evolution equations (\ref{Thetasub},
\ref{dtdtTheta}) for (the derivatives of) $\Theta$ form a separate
subsystem with the structure of the wave equation. Concerning the
remaining equations (\ref{Zsub}, \ref{dtdtz}), one can express
them in terms of the quantities
\begin{eqnarray}\nonumber
    Z_{ki} &\equiv& \partial_k Z_i + ~C_{ik} +~\gamma^{rs}~[~C_{ikrs}
    \\ \label{Zki} &+& ~(\zeta-1)~C_{rksi}+ ~(\zeta+1)~C_{risk}~]\,,
\end{eqnarray}
so that they read
\begin{eqnarray}
  1/\alpha^2~\partial_t~(\partial_t Z_i) - \partial^k~ [~Z_{ki}~]
  &=& \cdots\\
    \partial_t~ Z_{ki} - \partial_k~ [~\partial_t Z_i~] &=& \cdots
    \label{dtdtzki}\,,
\end{eqnarray}
and we get again the structure of the wave equation.

It follows that the principal part of the subsidiary system
(\ref{Asub} - \ref{dtdtz}) can be put in symmetric hyperbolic
form. The characteristic speeds are either zero or the light
speed. A simple energy estimate is provided by
\begin{eqnarray}\nonumber
    \mathbb{E} \equiv 1/\alpha^2~[~(\partial_t \Theta)^2 &+& \gamma^{ij}~
    (\partial_t Z_i)(\partial_t Z_j)~] \\
    +~(\partial_k \Theta)(\partial^k \Theta) &+& Z_{ij}Z^{ij}\,.
\label{estimate}
\end{eqnarray}

We are now in position to take the second step in the
constraint-preserving boundary conditions programme. We will
impose the vanishing of all the incoming modes of (\ref{Asub} -
\ref{dtdtz}) at the boundaries, that is
\begin{eqnarray}
  1/\alpha~\partial_t~\Theta + n^k~\partial_k~\Theta &=& 0
  \label{Thetabound}\\  \label{Zboun}
  1/\alpha~\partial_t~Z_i + n^k Z_{ki}  &=& 0\,,
\end{eqnarray}
where $\vec{n}$ stands here for the outwards-pointing unit normal
to the boundary surface.

Equations (\ref{Thetabound}, \ref{Zboun}) meet the two
requirements we were looking for:
\begin{itemize}
    \item They provide maximally-dissipative algebraic boundary
conditions for the subsidiary system (\ref{Asub} -\ref{dtdtz}). In
this way, no constraint-violating modes are allowed to enter
across the selected boundary.
    \item They will provide, as we will see in what follows, four
boundary conditions of the Sommerfeld type for the evolution
system (\ref{dtgamma}-\ref{dtZ}), which can be consistently
imposed in order to obtain true solutions of Einstein's field
equations. Notice that the extra terms in the definition
(\ref{Zki}) consist in ordering constraints, which would not
appear in a second order in space formulation.
\end{itemize}

\subsubsection{Boundary conditions implementation.}

The third step in the programme is to use the resulting values
($\Theta^{(boun)}$, $Z_i^{(boun)})$, as computed from
(\ref{Thetabound}, \ref{Zboun}), in order to obtain four of the
main system's incoming fields at the boundary. This process is not
free from ambiguities, like the choice of a suitable basis for the
dynamical fields.

For a symmetric hyperbolic evolution system, one could find a
(positive definite) quadratic form which would provide a metric
for the space of dynamical fields. The natural choice would be
then to build an orthogonal basis of dynamical fields containing
both $\Theta$ and $Z_i$ (or some equivalent combinations).
Imposing boundary conditions would then consist in prescribing the
values (\ref{Thetabound}, \ref{Zboun}) for these fields, while
leaving the remaining ones unchanged.

But the evolution system (\ref{dtgamma}-\ref{dtZ}) is not
symmetric hyperbolic. This means that we do not have a unique
prescription for imposing the boundary conditions, as far as we
have many ways of selecting an appropriate set of dynamical fields
at the boundary. A convenient starting point in this case is to
replace the original basis
\begin{equation}\label{Zbasis}
    (\Theta\,,~K_{ij}\,,~Z_i\,,~D_{kij}\,,~A_i)
\end{equation}
by  one which is more adapted to the characteristic decomposition
at the boundary, namely
\begin{equation}\label{Vbasis}
    (\Theta\,,~\tilde{K}_{ij}\,,~Z_i\,,
    ~D_{\bot ij}\,,~\tilde{D}_{n\bot\bot}\,,~V_i\,,~D_i\,,~A_i)\,,
\end{equation}
where the symbol $\bot$ replacing an index means the projection
orthogonal to $\vec{n}$. We have noted as $\tilde{D}_{n\bot\bot}$
the traceless part of $D_{n\bot\bot}$ and
\begin{equation}\label{tildeK}
      \tilde{K}_{ij} \equiv K_{ij}-\frac{\Theta}{2}~\gamma_{ij}~.
\end{equation}
Notice that the quantities $D_{nn\bot}, ~tr(D_{n\bot\bot})$ do not
appear explicitly in the new basis. These components must be
computed instead from $(Z+V)_i$ and the $D_{\bot ij}$ components.
Allowing for the definition (\ref{defVk}), we actually get
\begin{eqnarray}\label{DtZperp}
    D_{nn\bot} &=& D_\bot
    -h^{rs}\,D_{rs\bot} - (Z+V)_\bot \qquad\\
\label{DtZn}
   h^{rs}D_{nrs} &=&
    h^{rs}\,D_{rsn} + (Z+ V)_n\,,
\end{eqnarray}
where $h^{rs}$ stands for the (inverse) metric on the boundary
surface, namely
\begin{equation}\label{2metric}
h^{rs} \equiv \gamma^{rs}-n^rn^s~.
\end{equation}

The new basis (\ref{Vbasis}) has been chosen in such a way that,
as we can easily verify, the values of ($\Theta$, $Z_i$) appear in
only eight eigenfields (four characteristic cones), namely:
\begin{eqnarray}\label{E-bound}
    E^{\pm} &=& \Theta \pm V^n\,, \\
\nonumber
  L_{n\bot}^{\pm} &=& \tilde{K}_{n\bot}
  \pm [~\frac{1}{2}\;(A_\bot+D_\bot- 2 ~Z_\bot)
    \\ \qquad  &~& -~{\frac{\zeta+1}{2}}\,D_{\bot nn}
    +~{\frac{\zeta-1}{2}}\, h^{rs}\,D_{rs\bot}~]\,,
    \label{L-nAbound}\\ \nonumber
\label{L-trbound}
  L^{\pm}&=&
 h^{rs}\tilde{K}_{rs} \pm [~Z_n
  - \zeta ~h^{rs}D_{rsn}~]\,,
\end{eqnarray}
where we have noted
\begin{equation}\label{Ltilde}
     L^{\pm} \equiv  h^{rs}L_{rs}^{\pm} - E^{\pm}\,.
\end{equation}

In order to set up the required four boundary conditions, we will
simply replace the original values for $(\Theta~,Z_i)$ by
$(\Theta^{(bound)}~,Z^{(bound)}_i)$, while leaving the other
fields in the basis (\ref{Vbasis}) unchanged. To be more specific:
\begin{itemize}
    \item The original values for $(\Theta~,Z_i)$ are replaced by
$(\Theta^{(bound)}~,Z^{(bound)}_i)$, as computed from
(\ref{Thetabound}, \ref{Zboun}), respectively. This amounts,
modulo some linear combinations with tangent fields (transverse
derivatives), to prescribe the first term in $E^\pm$ and the
second terms in $(L_{n\bot}^{\pm},L^\pm)$.
    \item The values of their 'counterpart' fields
$(V_n\,,~\tilde{K}_{n\bot}\,,~h^{rs}\tilde{K}_{rs})$ are not
changed by the boundary conditions.
\end{itemize}

It is clear then that the values of the four characteristic cones
(\ref{E-bound},~\ref{L-nAbound},~\ref{Ltilde}) have been
prescribed in such a way that the four equations
(\ref{Thetabound}, \ref{Zboun}) hold true at the selected
boundary.

A further source of ambiguity comes from the prescription of the
remaining incoming eigenfields (the gauge and the transverse
traceless ones). We will use here a convenient generalization of
the maximally dissipative boundary conditions, namely
\begin{equation}\label{G-bound}
    \partial_t~G^- = 0\,, \qquad
    \partial_t~[~L_{\bot\bot}^{-}
    -\frac{1}{2}~(h^{rs}\,L_{rs}^{-})~\gamma_{\bot\bot}~] = 0\,,
\end{equation}
although we are aware that more sophisticated choices could be
required in physical applications.

\section{Constraints stability}

The final step in the proposed programme is to check the stability
of the constraint-preserving boundary conditions
(\ref{Thetabound}, \ref{Zboun}, \ref{G-bound}).

Notice however that the main evolution system is just strongly
hyperbolic, but not symmetric hyperbolic (at least not in the
generic case~\cite{Z4}). This means that the Majda-Osher
theory~\cite{MO75} can not be directly applied, and the same is
true for the Secchi theorems~\cite{Secchi96}. This is why we will
check the stability of (\ref{Thetabound}, \ref{Zboun},
\ref{G-bound}) by other methods, both at the theoretical and the
numerical level.

From the theoretical point of view, the well-known Fourier-Laplace
method~\cite{KL89} could provide necessary conditions for
stability~\cite{CS03}. We will prefer here a simpler approach, by
analyzing the system of equations verified by the dynamical fields
at the boundary. We will call it the modified system in order to
distinguish it from the original evolution system, which is being
used at the interior points. We will see that this approach
provides some insight about the behavior at the boundary points.
The drawback is that boundary points form just the outermost layer
of the computational domain. It follows that the modified system
analysis has to be considered at this stage just as an heuristic
approach, so that the stability of the boundary conditions must be
confirmed by other means. More details are provided in
Ref.~\cite{Foundations}.

\subsection{The modified system approach}

For the sake of clarity, let us focus first on the subset of
dynamical fields spanned by $(\Theta,~V_i)$. As stated in the
previous Section, the boundary conditions are not affecting any of
the $V_i$ components. This means that the boundary values of $V_i$
verify the main evolution system equations, namely
\begin{equation}
    1/\alpha~\partial_t~V_i + \partial_i~\Theta
    = \cdots  \label{dtV}
\end{equation}

The original equation (\ref{dtTheta}) for $\Theta$, however, no
longer holds at the boundary, where one is imposing instead the
advection equation (\ref{Thetabound}). This means that, even at
the continuum level, the evolution system is being modified at the
boundaries. The modified system for the subset of dynamical fields
($\Theta$,~$Z_i$) is given by (\ref{Thetabound}, \ref{dtV}).

The modified subsystem (\ref{Thetabound}, \ref{dtV}) has real
non-negative characteristic speeds along any direction $\vec{r}$,
oblique to $\vec{n}$. They are actually
\begin{equation}\label{speeds_energy}
        \{~0,~\alpha~(\vec{n}\cdot \vec{r})~\}\,.
\end{equation}
It follows that (the principal part of) the modified subsystem
(\ref{Thetabound}, \ref{dtV}) can be interpreted on physical
grounds as describing the outwards propagation of both $\Theta$
and $V_i$ at the boundary.

We can push one step further our analysis by considering the
particular case in which $\vec{r}\,$ is tangent to the boundary,
that is orthogonal to $\vec{n}$. In this case the speeds
(\ref{speeds_energy}) are fully degenerate, and a non-diagonal
coupling term remains in (\ref{dtV}), so that the modified
subsystem is just weakly hyperbolic. This has some relevant
consequences. Let us assume for instance that $\vec{n}$ is aligned
with the $x$ coordinate axis and that we get a static profile for
$\Theta$ of the form
\begin{equation}\label{Thetaprof}
    \Theta = g(y,z)~,
\end{equation}
which trivially satisfies equation (\ref{Thetabound}). The
derivative coupling in (\ref{dtV}) allows then modes in the $V_y$
and $V_z$ components which grow in time in a linear way. These
linearly growing modes will actually show up in numerical tests,
as we will see below.

The analysis of the full modified system can be simplified by
writing down (the principal part of) the modified evolution
equations for the combinations corresponding to incoming modes of
the original system. Allowing for (\ref{Thetabound}, \ref{Zboun}),
we have
\begin{eqnarray}\label{Emod}
  1/\alpha~\partial_t~E^- &=& 0 \\
  \nonumber
  1/\alpha~\partial_t~L_{n\bot}^{-} &=&
   h^{rs}\partial_r[~D_{ns\bot}-D_{sn\bot}
     +\frac{\zeta-1}{2}~\tilde{K}_{s\bot}] \\
\nonumber  &-&~\partial_\bot\,[~\frac{\zeta+1}{2}~\tilde{K}_{nn}
   -\frac{f+1}{2}~tr\,\tilde{K}+A_n\qquad\\
 &~& \qquad +~V_n -\frac{(3-2m)f+1}{4}~\Theta~] \\
\nonumber
  1/\alpha~\partial_t~L^{-} &=&
  -h^{rs}\partial_r\,[~D_{nns} - D_{snn}
  + \zeta\,\tilde{K}_{ns}\\
  &~& \qquad +~ A_s + V_s~]\,,
  \label{Lmod}
\end{eqnarray}
plus the trivial evolution equations (\ref{G-bound}) for the gauge
and the transverse traceless incoming modes.

Notice that only derivatives tangent to the boundary appear on the
modified system equations (\ref{Emod} - \ref{Lmod}) for the
incoming modes. This means that all the characteristic speeds
along the longitudinal direction $\vec{n}$ are real and
non-negative: they are actually
\begin{equation}\label{charmod}
    {0,~\alpha,~\alpha\sqrt{f}}\,.
\end{equation}
The corresponding eigenvectors are either standing fields ($v=0$):
\begin{equation}\label{modstanding}
    A_\perp\,,~~D_{\perp ij}\,,~~A_k-f D_k+f m
    V_k\,,~~E^-\,,~~L^-_{ij}\,,~~G^-
\end{equation}
or outgoing fields ($v=\alpha,~\alpha\sqrt{f}$):
\begin{equation}\label{modoutgoing}
    \Theta\,,~~Z_i\,,~~\tilde{L}^+_{\bot\bot}\,,~~G^+.
\end{equation}
These fields span the whole dynamical space; the modified system
is then strongly hyperbolic along the direction $\vec{n}$ normal
to the boundary.

Computing the characteristic speeds along a generic direction
$\vec{r}$, oblique to $\vec{n}$, and for an arbitrary value of the
ordering parameter, is a much harder task, even using an algebraic
computing program. We have just checked the particular cases
\begin{equation}\label{zetachoices}
    \zeta = 0\,,~\pm 1
\end{equation}
and we have found that the modified system is at least weakly
hyperbolic (real characteristic speeds) only in the $\zeta=0$
case. This suggests that the $\zeta=0$ case, corresponding to a
symmetric ordering of the space derivatives, could be free of
boundary instabilities, as we will confirm below.

\subsection{The robust stability test}

The robust stability numerical test~\cite{Mexico1} amounts to
consider small perturbations of Minkowski space-time which are
generated by taking random initial data for every dynamical field
in the system. The level of the random noise must be small enough
to make sure that we will keep in the linear regime even for
hundreds of crossing times (the time that a light ray will take to
cross the longest way along the numerical domain). We are taking
advantage in this way of the peculiar nature of the Einstein's
equations, where the principal part is quasilinear and the
non-principal (source) terms are quadratic in the dynamical
fields. Checking the linear regime of Einstein's equations amounts
then to test the behavior of their principal part.

This test has been previously used~\cite{Z48} for to check
numerically the stability properties of the Z4 evolution system
for interior points. In order to avoid boundary effects, the grid
had the topology of a three-torus, with periodic boundaries along
every axis. We will now open the $x$ faces and impose the
constraint-preserving boundary conditions there, while keeping
periodic boundary conditions along the other two axes.

\begin{figure}[t]
\begin{center}
\epsfxsize=8cm
\epsfbox{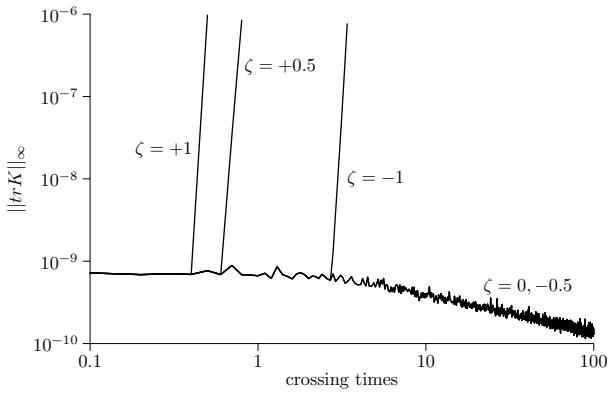}
\end{center}
\caption{$L_{\infty}$ norm of $trK$ for different values of the
ordering parameter $\zeta$ and constraint-preserving boundary
conditions along one single direction (periodic boundaries along
the other two). The values of $\zeta$ are shown with an interval
of $0.5$ for the sake of clarity, although the survey has been
made with a finer interval of $0.1$. The stable range is given by
$\zeta$ values in the interval $[-0.5,0]$. The decreasing of the
norm in the stable regions is due to the dissipative boundary
conditions (\ref{G-bound}) for the gauge modes}\label{1face}
\end{figure}

We show in Fig.~\ref{1face} the $L_{\infty}$ norm of $trK$ for
different values of the ordering parameter $\zeta$. A spacing
$\Delta\zeta=0.5$ is used in the plot for the sake of clarity,
although the numerical survey has been made with a finer spacing
of $\Delta\zeta=0.1$. Our results show that the
constraint-preserving boundary conditions (\ref{Thetabound},
\ref{Zboun}) are stable if and only if $\zeta\,$ is in the range
$[-0.5,0]$. The behavior is the same in all the stable regions:
the different values of $\zeta$ just determine when the
instabilities (if any) will appear.

\begin{figure}[t]
\begin{center}
\epsfxsize=8cm
\epsfbox{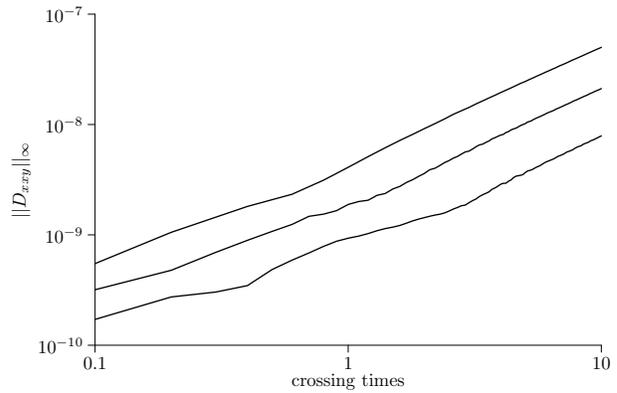}
\end{center}
\caption{Same as Fig.~\ref{1face}, but now for the $L_{\infty}$
norm of $D_{xxy}$ and $\zeta=0$. We plot here the results
corresponding to three different resolutions, focusing on the
first $10$ crossing times. Notice that this is a logarithmic plot,
so that the resolution-independent linear growing with unit slope
corresponds actually to a linearly growing dynamical
mode.}\label{1faceD}
\end{figure}

Notice that the robust stability analysis predicted the arising of
linearly growing modes related to the transverse $V_i$ components.
In terms of the original basis, one can expect to see these modes
in the quantities $D_{xxy}$, $D_{xxz}$, which are derived from
$V_x$, $V_z$ by the relationships (\ref{DtZperp}, \ref{DtZn}),
respectively. We can see for instance in Fig.~\ref{1faceD} a
growing linear mode in the $L_{\infty}$ norm of $D_{xxy}$. This
confirms that the modified system analysis can be useful to
anticipate the behavior of the boundary conditions under the
robust stability test. Further evidence in this direction is
provided in the Appendix B, where maximally dissipative boundary
conditions are considered.

The robust stability test is also useful for checking the
constraint-preserving character of the proposed boundary
conditions (\ref{Thetabound}, \ref{Zboun}). As far as true
Einstein's solutions are actually recovered by setting both
$\Theta$ and $Z_i$ to zero, the values of these quantities can be
considered to be good indicators of constraint violations. We can
monitor the norm of these quantities to check whether constraint
violations are being injected into the computational domain
through the open boundaries.

\begin{figure}[h]
\begin{center}
\epsfxsize=8cm \epsfbox{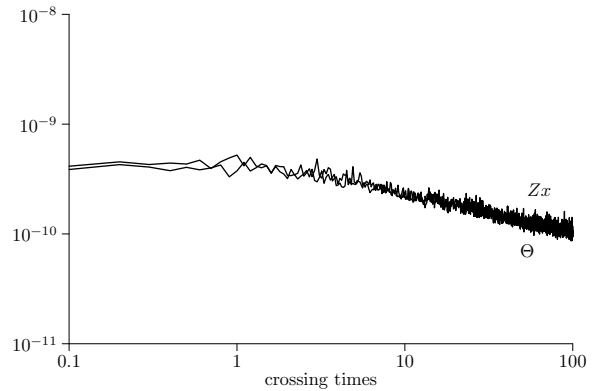}
\end{center}
\caption{Same as in the previous figures, but now the norms of
$\Theta$ and $Z_x$ are plotted in order to monitor constraint
violations. No growth can be seen, confirming that constraint
violations are not being injected through the
boundary.}\label{Zradevol}
\end{figure}

We can see in Fig.~\ref{Zradevol} that this is actually not the
case. The values of $\Theta$ and $Z_i$ are not growing at all,
contrary to what happens to the $D_{xxy}$ components, as seen in
Fig.~\ref{1faceD}. Moreover, their norm is diminishing: we can
understand this decreasing by noticing that the boundary
conditions (\ref{Thetabound}, \ref{Zboun}) are, modulo some
coupling with ordering constraints, advection equations. This
means that the values of (\ref{Thetabound}, \ref{Zboun}) are just
flowing out of the computational domain through the open boundary.

\section{Discussion and outlook}

One can wonder about wether the multi-dimensional character of the
problem is lost when applying boundary conditions to just one
face. This is not the case: the $x=constant$ boundary surfaces are
actually surfaces, not just points, so that oblique modes are
still present and can lead to instabilities, as it actually occurs
when $\zeta=\pm 1$. The only thing we are avoiding in this way is
to apply constraint-preserving boundary conditions to corner
points (which are assigned here to the $y$ and $z$ boundary
surfaces, where periodic boundary conditions are applied). If we
try to apply conditions (\ref{Thetabound} - \ref{Zboun}) along
every space direction, then instabilities appear even for the
$\zeta=0$ choice of the ordering parameter.

One could argue as well that the difficulties with corners and
edges could be related with the fact that the main evolution
system has just a strongly hyperbolic, but not symmetric
hyperbolic, principal part. This is not the case, as it is shown
in Appendix B, where boundary conditions of the maximally
dissipative type are successfully implemented and applied to all
the boundary points of a cartesian-like numerical grid, including
corners and edges.

Our results are very similar to those of Ref.~\cite{SW03}, where
the same test is applied to reflection boundary condition: a
stable linear-growing mode is detected, which becomes unstable
only when the boundary conditions are applied also to the other
faces, so that the numerical grid gets corners and edges. Our work
can be then understood as an extension (in a different formalism)
of the results of Ref.~\cite{SW03} to boundary conditions of the
Sommerfeld type.

In our opinion, the main problem at corner points comes from the
inconsistency inherent to the choice of a (unique) normal
direction there. Different faces get different normal vectors, but
corner points belong to two different faces at the same time. This
is not just a theoretical caveat: corners and edges pose a real
problem in practical applications, where more work should be done
along any of the following lines:
\begin{itemize}
    \item Devising an specific treatment for corner points. The
correct implementation of constraint-preserving boundary
conditions in the presence of corners is still a unsolved issue.
For symmetric hyperbolic evolution systems, using finite
difference operators satisfying the summation by parts rule with
respect to a diagonal scalar product leads to stable schemes with
maximally dissipative boundary conditions~\cite{CLRST04, CN04}. In
our case, with a strongly hyperbolic evolution system, we have got
the same result, as presented in Appendix B.

But these results do not extend to constraint-preserving boundary
conditions. A major difficulty is that compatibility conditions
between boundary data at adjacent faces need to be satisfied if
one wishes to obtain smooth solutions. Necessary conditions for
continuous solutions have been derived in Ref.~\cite{CPRST03} for
a symmetric hyperbolic evolution system. But more conditions are
needed in order to obtain smooth solutions. Compatibility issues
are also present at corner points between initial and boundary
data (see for instance Chapter 9 in Ref.~\cite{GKO95}).
    \item Building numerical grids with smooth boundaries (without
corners and edges), so that the constraint-preserving boundary
conditions (\ref{Thetabound} - \ref{Zboun}) can be applied
consistently in an stable way, as we have confirmed numerically by
means of the robust stability test. The construction of
'\,Multi-patch' numerical schemes, which would allow for smooth
boundaries, has become a major research topic in Numerical
Relativity. See for instance Refs.~\cite{Thorn04, CN04prep}.
\end{itemize}

\renewcommand{\theequation}{A.\arabic{equation}}
\setcounter{equation}{0}
\section*{Appendix A: Characteristic decomposition of
the Z4 first order system}

Let us consider the propagation of perturbations with wavefront
surfaces given by the unit (normal) vector $n_i$, we can write the
principal part of the Z4 first order system in matrix form
\begin{equation}\label{B_dtu}
  \frac{1}{\alpha}~\partial_t~\mathbf{u}
  ~+~ \mathbf{M} ~n^k\partial_k ~\mathbf{u} = ...~,
\end{equation}
where $\mathbf{u}$ is the full array of dynamical fields
(\ref{uvector}). Notice that derivatives tangent to the wavefront
surface play no role here.

A straightforward analysis of the characteristic matrix $\mathbf
{M}$ provides the following list of eigenfields~\cite{Z48}:

\begin{itemize}
    \item Standing eigenfields (zero eigenvalues)
\begin{equation}\label{B_EF0}
 \alpha\,,~~ \gamma_{ij}\,,~~A_\perp\,,~~
 D_{\perp ij}\,,~~A_k-f D_k+f m V_k\,,
\end{equation}
where the symbol $\perp$ replacing an index means the projection
orthogonal to $n_i$
\begin{equation}\label{B_DAij}
 D_{\perp ij} \equiv D_{kij} - n_k n^r D_{rij}.
\end{equation}

    \item Light-cone eigenfields (eigenvalues $\pm \alpha $)
\begin{eqnarray}\label{B_EFLij}
 {L^{\pm}}_{ij} &\equiv& [K_{ij}-n_i n_j ~{\tr}K ]
\nonumber \\
  &~& \pm~[{\lambda^n}_{ij} - n_i n_j~{\tr}\lambda^n]
\\
\label{B_EFL}
  E^{\pm} &\equiv& \Theta \pm V^n~,
\end{eqnarray}
where the symbol $n$ replacing the index means the contraction
with $n_i$
\begin{equation}
 \lambda^n_{ij}~\equiv~n_k~\lambda^k_{ij}\qquad V^n~\equiv~n_k~V^k.
\end{equation}

    \item Gauge eigenfields (eigenvalues $\pm \alpha \sqrt{f}$)
\begin{equation}\label{B_EFG}
 G^{\pm} \equiv \sqrt{f} \left[~{\tr}K -\mu\,\Theta ~\right]
  \pm \left[~A^n + (2-\mu)\,V^n ~\right]
\end{equation}
where we have noted for short
\begin{equation}\label{muparam}
    \mu \equiv \frac{f m-2}{f-1}~.
\end{equation}
In the degenerate case (f=1), one must have $m=2$, so that the
value of $\mu$ is not fixed. The degeneracy allows for any
combination with (\ref{B_EFL}), as expected.
\end{itemize}

\renewcommand{\theequation}{B.\arabic{equation}}
\setcounter{equation}{0}
\section*{Appendix B: Maximally dissipative boundary conditions}

A convenient generalization of the maximally dissipative boundary
conditions can be implemented by just imposing the vanishing of
(the time derivatives of) all the incoming modes, that is
\begin{equation}
\label{maxdiss}
  \partial_t~E^{-}
  =~ \partial_t~L_{ij}^{-} =~ \partial_t~G^{-} = 0~.
\end{equation}

The principal part of the modified system is the much simpler that
in the constraint-preserving case. It is, by construction,
strongly hyperbolic along the direction $\vec{n}$ normal to the
boundary, with characteristic speeds given again by
(\ref{charmod}).

We will compute the characteristic speeds along a generic
direction $\vec{r}$, oblique to $\vec{n}$, where the vector
$\vec{r}$ is related with $\vec{n}$ by
\begin{equation}\label{omega}
    \vec{r} = ~\vec{n} ~cos\varphi + \vec{s} ~sin\varphi\,,
\end{equation}
and we have taken
\begin{equation}\label{omega_cond}
    \vec{n}^{~2} = \vec{s}^{~2} = 1\,, \qquad \vec{n} \cdot \vec{s} = 0\,.
\end{equation}
The hyperbolicity requirement amounts to demand that all the
resulting characteristic speeds be real for any value of the angle
$\varphi$.

The trivial equations (\ref{maxdiss}) provide $7$ (remember that
$L_{ij}^{~\pm}$ is traceless) standing eigenfields (zero
characteristic speed) of the modified system. Another set of $17$
standing eigenfields is given by
\begin{equation}\label{ADV}
 A_p~, \qquad D_{p\,ij}~, \qquad A_k-f D_k+fm\, V_k~,
\end{equation}
where $\vec{p}~$ is the direction orthogonal to both vectors
$\vec{n}$ and $\vec{s}$.

The remaining $14$ dynamical fields can be grouped into the
following sectors:
\begin{itemize}
    \item \textbf{Energy sector} \{$E^+,~V_s$\}. The corresponding
evolution equations are (principal part only):
\begin{eqnarray}\label{bound_dV} \nonumber
    \frac{1}{\alpha}~\partial_t~V_s &=&
    - sin\varphi~\partial_s ~\Theta =
    - \frac{1}{2}~sin\varphi~\partial_r~E^+ \cdots\\ \nonumber
    \frac{1}{\alpha}~\partial_t~E^+ &=&
    - \partial_r[~V_r + \Theta~cos\varphi~] \\ \nonumber
    &=& - \partial_r[~E^+ cos\varphi + V_s ~sin\varphi+\cdots~]\,,
\label{bound_dE}\end{eqnarray} where the dots stand for coupling
terms with the standing eigenfields (which are irrelevant for the
eigenvalues calculation). It follows that the characteristic
speeds are given by the solutions of the algebraic equation
    \begin{equation}\label{lambdaE}
    \lambda(\lambda-\alpha~cos\varphi) =
    \frac{1}{2}\,\alpha^2~sin^2\varphi~,
    \end{equation}
    so that real characteristic speeds are obtained for every
    value of $\varphi$.
\item \textbf{Gauge sector} \{$G^+,~A_s$\}. The corresponding
evolution equations are (principal part only):
\begin{eqnarray}\nonumber
    \frac{1}{\alpha}~\partial_t A_s &=&
    - sin\varphi~\partial_s [~f(tr K - m~\Theta)] \\ \nonumber
    &=& - \frac{1}{2}\sqrt{f}\,sin\varphi\,
    \partial_r~G^+ +~\cdots \label{bound_dA}\\ \nonumber
    \frac{1}{\alpha}~\partial_t~G^+ &=&
    - \partial_r[~\sqrt{f}\,A_r + f~cos\varphi~tr K~] \\ \nonumber
    &=& - \sqrt{f}\,\partial_r[~G^+~cos\varphi~
    + A_s ~sin\varphi ~+ \cdots]\label{bound_dG}
    \end{eqnarray}
    where the dots stand for coupling terms with the previous
    sectors. It follows that the characteristic speeds are
    given by the solutions of the algebraic equation
    \begin{equation}\label{lambdaG}
    \lambda(\lambda-\alpha~\sqrt{f}\,cos\varphi) =
    \frac{1}{2}\,f~\alpha^2~sin^2\varphi~,
    \end{equation}
    so that, allowing for the positivity of the gauge parameter $f$,
    real characteristic speeds are obtained again for every value
    of $\varphi$.
\item \textbf{Metric sector} \{$L_{ij}^{~+},~D_{sij}$\}. The
corresponding evolution equations can be written as (principal
part only)
\begin{eqnarray}\nonumber
    \frac{1}{\alpha}&\partial_t& D_{sij} =
    - sin\varphi~\partial_s ~K_{ij}
    = - \frac{1}{2}\,sin\varphi\,
    \partial_r[~L_{ij}^{~+}+\cdots~] \label{bound_dD}\\
    \frac{1}{\alpha}&\partial_t&L_{ij}^{~+} =
    - \partial_r[~\lambda^r_{ij} + cos\varphi~K_{ij} + \cdots\\ \nonumber
    &~& - \frac{1+\zeta}{2}~(r_i K_{nj}+r_j K_{ni} -
    n_i K_{rj} - n_j K_{ri}) ~] \\ \nonumber
    &=& - \partial_r[~L_{ij}^{~+}~cos\varphi + D_{sij}~sin\varphi
    + \cdots\\ \nonumber
    &~& - \frac{1+\zeta}{2}~sin\varphi~(s_i K_{nj}+s_j K_{ni} -
    n_i K_{sj} - n_j K_{si}) \\ \nonumber
    &~& - \frac{1+\zeta}{2}~sin\varphi~(D_{ijs}+D_{jis} -
    s_i E_{j} - s_j E_{i})~]\,.
\label{bound_dL}\end{eqnarray}
    where the dots stand again for coupling terms with the previous sectors.
\end{itemize}

The evolution equation (\ref{bound_dL}) for these outgoing
'\,metric' fields contains (unless $\zeta = -1$) crossed coupling
terms that complicate the analysis. One gets three variants of the
same algebraic equation
\begin{eqnarray}
    \lambda(\lambda-\alpha~cos\varphi) &=&
    \frac{1}{2}~\alpha^2~sin^2\varphi \\
    \lambda(\lambda-\alpha~cos\varphi) &=&
    \frac{1}{2}~\alpha^2~sin^2\varphi ~[1-(\frac{1+\zeta}{2})^2]
    \qquad\\
    \lambda(\lambda-\alpha~cos\varphi) &=&
    \frac{1}{2}~\alpha^2~sin^2\varphi ~[1-(1+\zeta)^2]\,,
    \label{lambda_conds}
\end{eqnarray}
depending on the particular set of components considered (the last
two equations appear twice, so that one gets $10$ characteristic
speeds that complete the full set of $38$). The most restrictive
is the last one (\ref{lambda_conds}): it implies that we get
complex characteristic speeds for some values of $\varphi$ unless
\begin{equation}\label{zeta_cond}
    \zeta \leq 0\,,
\end{equation}
so that the standard ordering case ($\zeta=+1$) is excluded.

\begin{figure}[t]
\begin{center}
\epsfxsize=8cm
\epsfbox{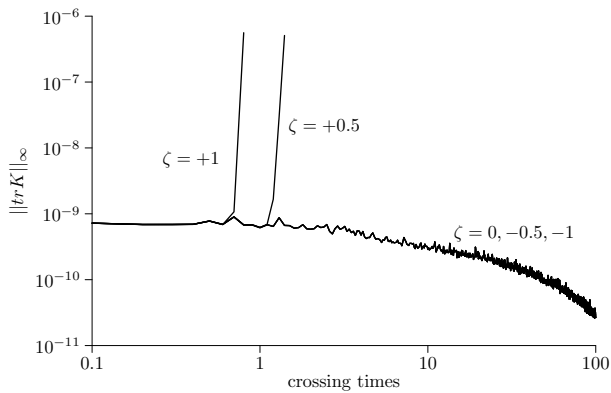}
\end{center}
\caption{Same as Fig.~\ref{1face}, but now with maximally
dissipative boundary conditions enforced along every axis. Again,
the values of $\zeta$ are shown with an interval of $0.5$ for the
sake of clarity, although the survey has been made with a finer
interval of $0.1$. It confirms that the $\zeta \leq 0$ choices are
stable, as expected from the modified system analysis. Notice
that, in the stable cases, the decrease after $100$ crossing times
is more than one order of magnitude greater than in
Fig.~\ref{1face}. This shows the effect of the maximally
dissipative boundary conditions: they are actually dissipating all
the dynamical fields.}\label{bound_strong}
\end{figure}

We can check out these results by using again the robust stability
test-bed. We will enforce the maximally dissipative conditions
(\ref{maxdiss}) along every axis in a cartesian-like numerical
grid, including corners and edges. We will survey the values of
the $\zeta$ parameter in the interval $[-1,1]$, with a spacing
$\Delta\zeta=0.1$.

\begin{figure}[t]
\begin{center}
\epsfxsize=8cm \epsfbox{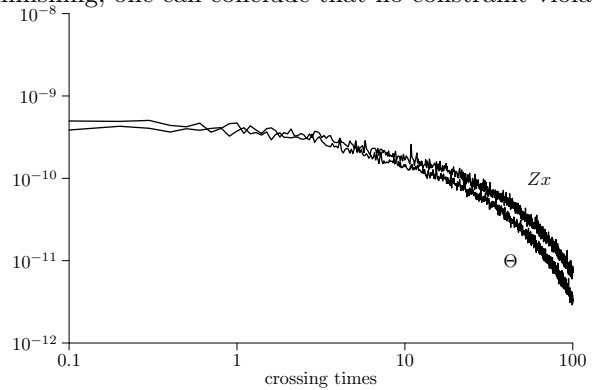}
\end{center}
\caption{Same as in the previous figure, but now the norms of
$\Theta$ and $Z_x$ are plotted in order to monitor constraint
violations. Their decay indicates that the constraint-violating
modes are diminishing, even much faster than in
Fig.~\ref{Zradevol}. But now the reason is a different one: the
boundary conditions are dissipating all the dynamical fields.
}\label{Zdisevol}
\end{figure}

We show in Fig.~\ref{bound_strong} the time evolution of the
maximum of the absolute value of $tr K$ (a spacing
$\Delta\zeta=0.5$ is used in the plot for the sake of clarity).
Our results show that the positive choices of the ordering
parameter are actually unstable, whereas the choices in the range
$[-1,0]$ are stable and behave in the same way. Notice that the
norm of $tr K$ in Fig.~\ref{bound_strong} is decreasing, in the
stable cases, much faster than in the constraint-preserving case
(Fig.~\ref{1face}). This can not be explained just by the fact
that boundary conditions are now being applied along the three
coordinate axes: the boundary conditions are actually dissipating
all the dynamical fields.

We show in Fig.~\ref{Zdisevol} the time evolution of the maximum
of the absolute value of both $\Theta$ and $Z_i$ for the symmetric
ordering ($\zeta=0$) case. As far as their values are diminishing,
one can conclude that no constraint-violating modes are being
produced at the boundaries. But notice that this is at the price
of dissipating all the dynamical fields. One can not conclude then
that maximally dissipative boundary conditions are
constraint-preserving: constraint-related fields are flowing out
through the boundaries in the same way as the remaining degrees of
freedom.

{\em Acknowledgements: This work has been supported by the EU
Programme 'Improving the Human Research Potential and the
Socio-Economic Knowledge Base' (Research Training Network Contract
HPRN-CT-2000-00137), by the Spanish Ministry of Science and
Education through the research project number FPA2004-03666 and by
a grant from the Department of Innovation and Energy of the
Balearic Islands Government.}

\bibliographystyle{prsty}

\end{document}